\documentclass[twocolumn,amsmath,amsfonts,showpacs,floatfix]{revtex4}
\usepackage{graphicx}% Include figure files
\usepackage{tikz}

\begin{document}
\title{Entanglement switching via the Kondo effect in triple quantum dots}

\author{S. B. Tooski$^{1,2,3}$, A. Ram\v sak$^{1,2}$,  R. \v Zitko$^{2}$, and B. R. Bu{\l}ka$^{3}$}
\affiliation{ $^{1}$ Faculty of Physics and Mathematics, University of Ljubljana, Jadranska 19, Ljubljana, Slovenia}
\affiliation{ $^{2}$ Jo$\check{z}$ef Stefan Institute, Jamova 39, Ljubljana, Slovenia}
\affiliation{ $^{3}$ Institute of Molecular Physics, Polish Academy of Science, ul. M. Smoluchowskiego 17, 60-179 Pozna{\'n}, Poland}

\date{\today}

\begin{abstract}
We consider a triple quantum dot system in a triangular geometry with one of the dots connected to metallic leads. Using Wilson's numerical renormalization group method, we investigate quantum entanglement and its relation to the thermodynamic and transport properties, in the regime where each of the dots is singly occupied on average, but with non-negligible charge fluctuations. It is shown that even in the regime of significant charge fluctuations the formation of  the Kondo singlets induces switching between separable and perfectly entangled states. The quantum phase transition between unentangled and entangled states is analyzed quantitatively and the corresponding phase diagram is explained by exactly solvable spin model.
 
\end{abstract}


\maketitle

\section{Introduction}

Creation of entangled states is essential for quantum computation and
communication  where most qubit operations cannot be performed through the manipulation of separable states \cite{horodeccy}.  Since the electron spin is a natural two-level system the production and manipulation of spin qubit pair entanglement attracted much attention \cite{Nielsen,Luczak,marcin}.
Two electron spin qubits, each localized in one of two adjacent
semiconductor quantum dots (QDs), can be coupled via the Heisenberg
exchange interaction due to virtual electron tunneling between the
dots \cite{Loss}. The description of electrons by the spin degrees of
freedom alone is a simplification valid when the electrons are
localized and the charge fluctuations are negligible. In general, in
any realistic solid-state device, spin entanglement is closely
connected to the orbital degrees of freedom of the carriers which can
be traced out if not measured \cite{Ramsak1}. Following the analysis
of the use of entangled electron spin pairs in solid-state structures
\cite{Burkard}, an intense activity was aimed toward understanding the
physical mechanisms that produce spin-entangled electrons in
mesoscopic conductors. The coherent manipulation of a single electron
spin in a QD and the controlled correlation of two spins located in
isolated dots have already  been demonstrated experimentally \cite{Hanson}. 

The interaction of qubit pairs with the environment is in general a
complicated many-body process and its understanding is essential for
experimental and theoretical solid-state qubits \cite{Eriksson}. Such
a system is often represented by a two-level system, or spin-1/2,
interacting with an otherwise homogeneous, and often one-dimensional
medium with gapless excitations. Some versions of this model are
equivalent to the Kondo model, motivating studies of ground state
entanglement of an impurity spin with the conduction electrons and the
role of the Kondo effect. This entanglement can be easily expressed
exactly in terms of the impurity magnetization
\cite{Eriksson,Ramsak2}. It was found that the Kondo effect plays a significant role in spin qubit double-quantum dot (DQD) \cite{Ramsak2}.
\begin{figure}[hb]
\centering
\includegraphics[width=0.3\textwidth]{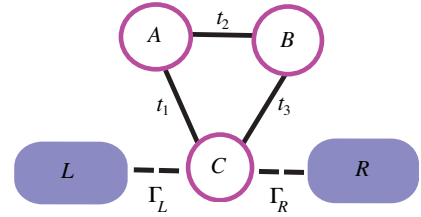}
\caption{Triple  quantum dot system attached to the leads.}
\label{fig1}
\end{figure}

Recently, the transport properties of triple quantum dots (TQD) in
various configurations have been investigated
\cite{Zitko,Wang,Mitchell,Chiappe,baruselli}. In a serial TQD,
different Kondo regimes can be sampled by measuring the conductance.
The transition from local-moment to molecular-orbital regime can be
observed in the evolution of correlation functions \cite{Zitko}. In a
system containing a central interacting dot C attached to the leads
and two non-interacting dots $A$ and $B$, there is a sizable splitting
of the Kondo resonance \cite{Wang} and a two-stage screening of the
magnetic moment was found. The inter-dot hopping introduces the
Kondo-assisted transport and may induce a quantum phase transition
\cite{Mitchell}.  The possibility of a robust underscreened Kondo
effect in TQD has also been recently discussed \cite{baruselli}.

In this work we concentrate on thermodynamic and transport properties and their relation to entanglement for three electrons confined in three adjacent dots forming a ring; one of the dots is attached to metallic leads, Fig.~\ref{fig1}. We study the case in which the dots are singly occupied on average. The Kondo effect can switch on the entanglement due to the interplay between the inter-dot spin-spin correlations and various Kondo-like ground states. We find conditions for entangled and unentangled states between the dots. The results differ from those in the DQD configurations in which the Kondo interaction reduces the entanglement between the DQD qubits \cite{Ramsak2}.

The outline of the paper is as follows. Sec. II introduces the model and numerical method. In Sec. III numerical results are presented and summarized in a phase diagram and effective model in Sec. IV.

\section{Model and method}

We  consider a solid-state qubit system built from a triangular TQD.
We model it by the three-impurity Anderson Hamiltonian. The
concurrence will be used as the measure of spin entanglement \cite{Ramsak2,Bennett}. We tune the parameters so that each dot is approximately singly occupied. The total spin is conserved in this system, thus the concurrence is related to spin-spin correlation functions.

The Hamiltonian has three parts: $H= H_{d}+H_{lead}+H_{t}$. The isolated dots are described by 
\begin{eqnarray}
\label{quad1}
H_{d}&=& \sum_{i}\epsilon_{i} d_{i}^\dag d_{i}+
U\sum_{i}  n_{i,\sigma}n_{i,\bar{\sigma}}+\\
&+&\sum_{i<j,\sigma}(t_{ij}d_{i \sigma}^\dag d_{j \sigma}+h.c.),
\label{hubbard}
\end{eqnarray}
where $n_{i,\sigma}=d_{i\sigma}^\dag d_{i\sigma}$ is the occupation of
dot $i\in\{A,B,C\}$ with spin $\sigma$. Only one energy level $\epsilon_{i}=\epsilon$ is taken into account in each quantum dot --  other levels are assumed to be much higher in energy. Parameter $U$ describes the Coulomb interaction between two electrons in the same orbital. We use compact notation for the hopping integrals: $t_1=t_{AC}$, $t_2=t_{AB}$, and $t_3=t_{BC}$, Fig.~\ref{fig1}. The dot C is coupled to the left $(L)$ and the right $(R)$ electrode,
\begin{eqnarray}
\label{quad2}
 \begin{split}
H_{t}= \sum_{\alpha, k,\sigma}(V_\alpha c_{\alpha k \sigma}^\dag d_{C \sigma}+h.c.),
\label{hubbard2}
\end{split}
\end{eqnarray}
where $V_\alpha$ is the tunnel matrix element between lead $\alpha$ and dot C. Electrons in the leads are described by a non-interacting Hamiltonian
\begin{eqnarray}
\label{quad3}
H_{lead}= \sum_{\alpha, k,\sigma}(\epsilon_{k} c_{\alpha k \sigma}^\dag c_{\alpha k\sigma}+h.c.),
\label{hubbard3}
\end{eqnarray}
where $\epsilon_{k}$ with  wavevector $k$ is single-electron energy. $d_{i \sigma}^\dag$ ($d_{i \sigma}$) and $c_{\alpha k \sigma}^\dag$ ($c_{\alpha k \sigma}$) are the creation (annihilation) operators in the dot and in the lead, respectively.
The leads have constant density of states $\rho=1/(2D)$, where D is the half-bandwidth of the conduction band. 
The hybridization strength is $\Gamma=\pi \rho \left(
|V_L|^2 + |V_R|^2 \right)$. Throughout this paper we assume $V_L=V_R$ and $t_1=t_3$.

To properly describe the Kondo effect \cite{Krishna-murthy,NRG,Hewson}, 
impurity problems need to be solved using nonperturbative methods, such as the numerical renormalization-group (NRG). The NRG \cite{Krishna-murthy,NRG} consists of a logarithmic discretization of the continuum of states of the conduction-band electrons, followed by a mapping to a one-dimensional chain Hamiltonian with exponentially decreasing hopping constants. 
As a consequence of the logarithmic discretization, the hopping along
the chain decreases exponentially, $t_{n}\approx\Lambda^{-n/2}$, where
$\Lambda$ is the discretization parameter and $n$ is the index of the
site in the chain. This provides the opportunity to diagonalize the
chain Hamiltonian iteratively and to keep only the states with the
lowest lying energy eigenvalues, since the energy scales are
separated. Knowing the energy eigenstates and eigenvalues, we can
calculate thermodynamical and dynamical quantities directly ({\it e.g.},
spectral functions using their Lehmann representation). The method is
reliable and rather accurate. Calculations in this work are performed
with a discretization parameter $\Lambda=2$, four values of the twist
parameter $z$, and the truncation energy cutoff of $E_\mathrm{cutoff}=10\omega_{N}$, where $\omega_{N}$ is the characteristic energy scale at the Nth NRG iteration. 

Since the NRG is a well-established numerical method with a relatively
broad literature, we do not give the details of the implementation
here but refer to Ref. \cite{Krishna-murthy,Costi,Bulla2,Hofstetter}.
In this work we use the NRG Ljubljana code \cite{NRG}.

\section{Results}
\subsection{Quantum entanglement}

Our goal is to analyse the entanglement of a qubit subsystem attached to external reservoir. A similar study of DQDs has shown that at low temperatures and low magnetic field, the Kondo effect acts as a source of entanglement destruction \cite{Ramsak1,Ramsak2}. On the contrary, in the present triple quantun dot system, the Kondo effect enables the entanglement, as shown in the following.

Quantum entanglement of two-qubits in a pure state
$|AB\rangle=\alpha_{\uparrow\uparrow}|\!\uparrow\uparrow\rangle+
\alpha_{\uparrow\downarrow}|\!\uparrow\downarrow\rangle+
\alpha_{\downarrow\uparrow}|\!\downarrow\uparrow\rangle+
\alpha_{\downarrow\downarrow}|\!\downarrow\downarrow\rangle$ can customarily be quantified by the Wootters concurrence \cite{Bennett} 
\begin{equation}
C_{AB}=2 |\alpha_{\uparrow\uparrow} \alpha_{\downarrow\downarrow} -\alpha_{\uparrow\downarrow} \alpha_{\downarrow\uparrow}|.
\end{equation}
Two qubits are fully entangled, $C_{AB}=1$, if they are in one of the
Bell states:
$|\!\uparrow\downarrow\rangle\pm|\!\downarrow\uparrow\rangle$ or
$|\!\uparrow\uparrow\rangle\pm|\!\downarrow\downarrow\rangle$. 

In TQD, the qubits are coupled to a fermionic bath, therefore the
two-qubit system AB is in a mixed state and appropriate generalization
for concurrence is given by the Wootters formula \cite{Bennett}.
Moreover, in the system considered there are also charge fluctuations,
thus the state may be outside the simple manifold of spin degrees of
freedom. Nevertheless, the concurrence can still be consistently
defined for such system. In the presence of axial spin symmetry, it can be given as \cite{Ramsak1,Ramsak2}
\begin{eqnarray}\label{cab2}
C_{AB}&=&\frac{\max(0,C_{\uparrow\downarrow},C_{||})}{P_{\uparrow\downarrow}+P_{||}},\\ \label{Cantiparallel}
C_{\uparrow\downarrow}&=&2|\left\langle S^{+}_{A} S^{-}_{B} \right\rangle| - 2\sqrt{\left\langle P^{\uparrow}_{A} P^{\uparrow}_{B} \right\rangle \left\langle P^{\downarrow}_{A} P^{\downarrow}_{B} \right\rangle },\\ \label{Cparallel}
C_{||}&=&2|\left\langle S^{+}_{A} S^{+}_{B} \right\rangle| - 2\sqrt{\left\langle P^{\uparrow}_{A} P^{\downarrow}_{B} \right\rangle \left\langle P^{\downarrow}_{A} P^{\uparrow}_{B} \right\rangle },\\ \label{Pantiparallel}
P_{\uparrow\downarrow}&=&\left\langle P^{\uparrow}_{A} P^{\downarrow}_{B} + P^{\downarrow}_{A} P^{\uparrow}_{B} \right\rangle,\\ \label{Pparallel}
P_{||}&=&\left\langle P^{\uparrow}_{A} P^{\uparrow}_{B} + P^{\downarrow}_{A} P^{\downarrow}_{B} \right\rangle,
\end{eqnarray}
where $S^{+}_{i}=(S^{-}_{i})^\dag=c_{i \uparrow}^\dag c_{i
\downarrow}$ is the electron spin raising operator for dot $i=A$, $B$
and $P^{\sigma}_{i}=n_{i \sigma}(1-n_{i \sigma})$ is the projection
operator onto the subspace where dot $i$ is singly occupied by one electron with the spin $\sigma$. $P_{||}$ and $P_{\uparrow\downarrow}$ are probabilities for the spins to be aligned in the same (parallel) and opposite (antiparallel) directions, respectively. This formalism is also applicable for the concurrence between other dots, {\it e.g.} $C_{CA}$ (concurrence between the dot C with the dots A) and $C_{CB}$ (concurrence between the dot C with the dots B). The expectation values can be easily obtained from the NRG.

\subsection{Thermodynamic quantities and transport}

In order to better understand the nature of the creation of the entanglement studied here we additionally calculate:

(1) The temperature-dependent impurity contribution to the impurity magnetic susceptibility \cite{Krishna-murthy}
\begin{eqnarray}\label{sus}
\chi_\mathrm{imp}(T)=\frac{(g \mu_{B})^2}{k_{B}T}\left(\left\langle S^{2}_{z}\right\rangle - \left\langle S^{2}_{z}\right\rangle_{0}\right),
\end{eqnarray}
where $S_{z}$ is the $z$ component of the total spin of the whole
system while $\left\langle ...\right\rangle$ means the thermodynamic
expectation values. The first expectation value refers to the system
with dots, while the second (with the subscript 0) refers to the
system without dots; $\mu_{B}$  is the Bohr magneton, $g$ is the
$g$-factor, and $k_{B}$ is the Boltzmann constant. It should be noted
that the combination $k_{B} T \chi_\mathrm{imp}(T)/(g \mu_{B})^2$  can be considered as an effective moment of the impurities.

(2) The impurity contribution to the entropy \cite{Krishna-murthy}
\begin{eqnarray}\label{entropy}
S_\mathrm{imp}(T)=\frac{(E-F)}{T}-\frac{(E-F)_{0}}{T},
\end{eqnarray}
where $E=\left\langle H\right\rangle=\mathrm{Tr}[H\exp(-H/k_{B}T)]$
and $F=-k_{B}T\ln \mathrm{Tr}[\exp(-H/k_{B}T)]$.

(3) Thermodynamic expectation values of various operators such as the on-site occupancy $\left\langle n_{i}\right\rangle$, local charge fluctuations $\left\langle (\delta n_{i})^2\right\rangle=\left\langle n^{2}_{i}\right\rangle-\left\langle n_{i}\right\rangle^{2}$, and spin-spin correlations $\langle \mathbf{S}_i\cdot\mathbf{S}_j\rangle$.

(4) The electronic transport between the leads through dot $C$, making use of the Meir-Wingreen formula \cite{Meir} 
\begin{eqnarray}\label{G2}
G(T)=G_{0}\pi\Gamma \int_{-\infty}^{+\infty}d\omega\left(-\frac{\partial f}{\partial \omega}\right)A_{C}(\omega,T),
\end{eqnarray}
where $G_{0}=2e^2/h$ is the conductance quantum, $f$ is the Fermi function, and $A_{C}(\omega,T)$ is the spectral function on the impurity $C$.

\subsection{Numerical analysis}

\begin{figure}[ht]
\includegraphics[width=.331\textwidth]{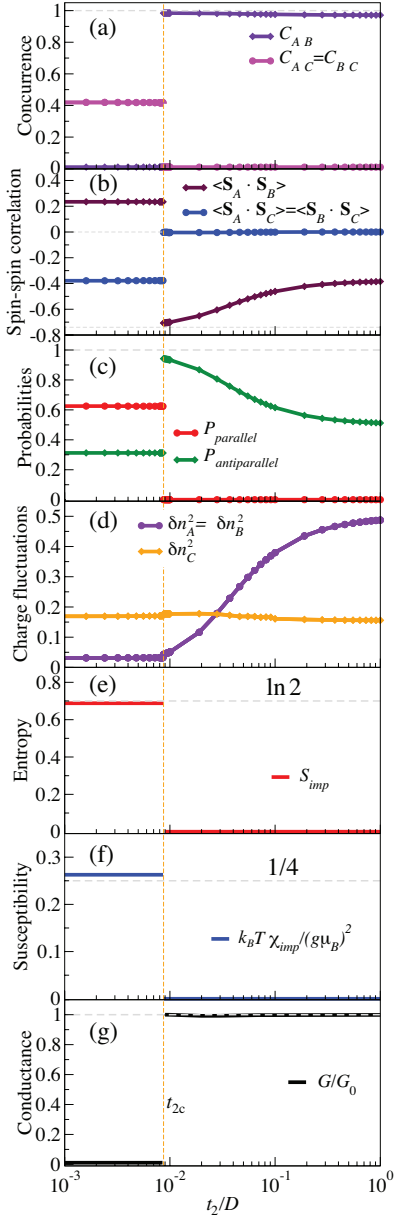}
\caption{Concurrence, spin-spin correlations, probabilities for
parallel and antiparallel spin configuration, charge fluctuations,
entropy $S_\mathrm{imp}/k_{B}$, susceptibility $k_B T \chi_\mathrm{imp}(T)/(g \mu_{B})^2$, and conductance at temperature $T=10^{-6}D$, as a function of $t_{2}/D$ for $U/D=0.1$, $\Gamma/D=0.01$, $\epsilon=-U/2$, $t_{1}/D=0.01$.}
\label{fig2}
\end{figure}

\begin{figure}[ht]
\includegraphics[width=.33\textwidth, clip]{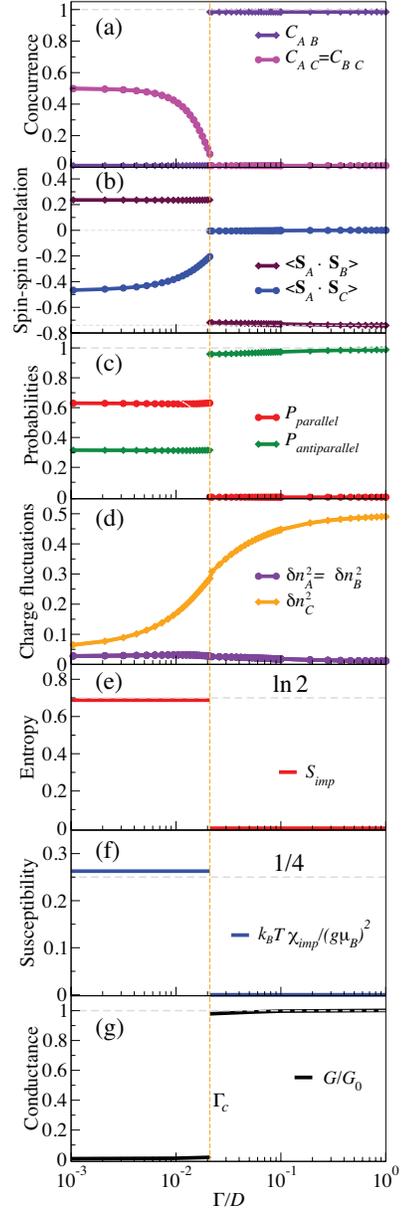}
\caption{Concurrence, spin-spin correlations, probabilities, and charge fluctuations at zero temperature as a function of $\Gamma/D$ for $t_{2}/t_1=0.5$; other parameters as in Fig.~\ref{fig2}.}
\label{fig3}
\end{figure}

%\red{}

The analysis of spin-spin correlation functions demonstrates the
singlet-triplet transition between the dots $A$ and $B$. The two
regimes are separated by a first order quantum phase transition (level
crossing) as a function of  $t_2/t_1$ which determines the ground state of the isolated TQD \cite{Wang,Mitchell,baruselli}. The same quantum phase transition has been previously found in a double quantum dot system modeled as a pure spin system \cite{vojta}. 
Entanglement properties of a qubit pair formed by spins on adjacent
quantum dots are closely related to the spin-spin correlations
\cite{Ramsak1} therefore  transitions between different spin
configurations play a crucial role. 

We first describe the strongly correlated regime with $U=10\Gamma$ at temperature  below the Kondo scale (essentially we are in the $T=0$ limit). Each of the dots is almost perfectly singly occupied. It should be noted that contrary to some double QD systems here exact single occupancy of identical dots  in general cannot be achieved for a common value of $\epsilon$ since the system is asymmetric and, furthermore, it is not particle-hole symmetric due to the
lack of bipartiteness of the lattice.
Expectation values $\left\langle ...\right\rangle$ in the concurrence formula Eq.~(\ref{cab2}) correspond to the thermal equilibrium of the system and consequently $\left\langle S^{+}_{A}\ S^{+}_{B}\right\rangle=0$. In vanishing magnetic field, the concurrence formula \eqref{cab2} simplifies further to
\begin{equation}
C_{AB}=\max\left\{0,-2\langle \mathbf{S}_A\cdot\mathbf{S}_B\rangle/(P_{\uparrow\downarrow}+P_{||})-1/2\right\}.
\end{equation}
Therefore, the concurrence is significant with increased spin-spin correlations in the range $-3/4\leq\langle \mathbf{S}_A\cdot\mathbf{S}_B\rangle\leq-1/4$.

In Fig.~\ref{fig2}(a) we show the concurrence between the adjacent
dots. As expected, there are two regimes of interest, depending on the inter-dot couplings, similar to the result of a recent analysis of thermal entanglement of isolated TQDS \cite{marcin}. On one hand, for $t_{2}>t_{2c}$ we get perfect entanglement between the dots $A$ and $B$, which are antiferromagnetically coupled, Fig.~\ref{fig2}(b). Critical interdot coupling $t_{2c}$ is equal to $t_1$ for decoupled quantum dots, $\Gamma=0$, and is slightly renormalized due to the coupling to the leads \cite{vojta,Wang,Mitchell}. For increasingly large coupling $t_2$, the spin-spin correlations diminish due to charge fluctuations, Fig.~\ref{fig2}(d), which reduce the probability for single occupation of the dots, Fig.~\ref{fig2}(c). However, the concurrence remains constant, $C_{AB}=1$. This means that two electrons would form a perfectly entangled qubit pair, if extracted one from the dot $A$ and the other from $B$. The dot $C$ is in the Kondo regime with electrons in the leads and it is completely decoupled from the other dots. Similar behavior, known as a dark-spin state, occurs in TQD ring in presence of an in-plane electric field due to decoupling of the spin in one of the dots \cite{Luczak}. On the other hand, for $t_{2}\lesssim t_{1}$, the concurrence is zero and ferromagnetic correlations between dots $A$ and $B$ dominate, forming an effective $S=1$ impurity at low temperatures which
undergoes partial Kondo screening, yielding a residual uncompensated spin-$1/2$.

In addition to the concurrence, spin and charge correlation
functions, abrupt changes also occur in thermodynamic properties. In
Fig.~\ref{fig2}(e) we show the entropy change from $S_\mathrm{imp}/k_B=\ln
2$ to zero and the susceptibility from $\chi_\mathrm{imp}=1/4$ to 0,
Fig.~\ref{fig2}(f). Finally, at the transition the linear conductance
changes from zero to $G=G_{0}$. This is expected due to the decoupling of the dot $C$ from the entangled pair $AB$ -- the conductance is unity as in the case of a single Anderson impurity.

The transition at $t_{2c}\sim t_1$ is explained by the level crossing
of the corresponding doublet eigenstates of isolated TQD
\cite{Wang, Mitchell,marcin,Luczak}.  However, for $t_{2}$ much
lower than $t_{1}$ full-entanglement between the dots $A$ and $B$ can
be restored by a sufficient increase of the effective Kondo coupling
of the dot $C$  to the leads, leaving the dots $A$ and $B$ in the
singlet state. In Fig.~\ref{fig3}(a), we show how the concurrence
$C_{AB}$  abruptly changes from zero to unity with increasing dot-lead
hybridisation $\Gamma$. This is an alternative way of inducing the
same quantum phase transition, as evident from the  entropy,
susceptibility and the conductance which all exhibit behavior
analogous to the results presented in Fig.~\ref{fig2}. The main
difference is in the fact that dots $A$ and $B$ remain consistently
decoupled from the rest of the system for the whole range of $\Gamma$, as is seen from the spin-spin correlations signaling triplet or singlet state for $\Gamma<\Gamma_c$ and $\Gamma>\Gamma_c$, respectively, while $\langle \mathbf{S}_A\cdot\mathbf{S}_C\rangle\sim0$ due to the increase of
$\delta n_C^2$ with increasing $\Gamma/D$. Charge fluctuations $\delta n_A^2$ remain low, as expected.

\begin{figure}[bt]
\includegraphics[width=0.5\textwidth, clip]{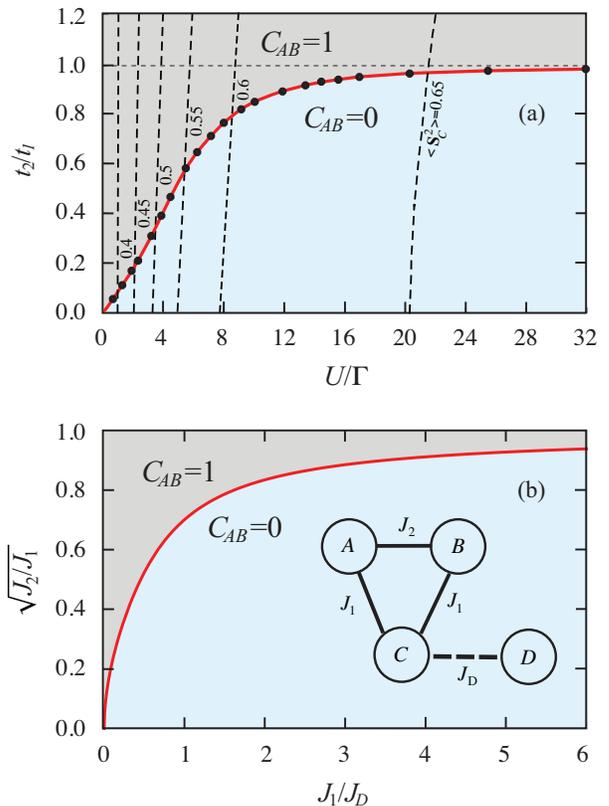}
\caption{(a) Phase diagram in the $(U/\Gamma,t_2/t_1)$ plane for $t_{1}/D=0.01$ and fixed $U/D=0.1$. Full red line (connecting calculated $t_{2c}/t_1$, bullets) separates $C_{AB}=1$ (the Kondo phase) and $C_{AB}=0$ regions between two ground state configurations. Dashed lines represent constant  values of local moment $\langle\mathbf{S}_C^2\rangle$. (b) Phase diagram in the $(J_1/J_{D},\sqrt{J_2/J_1})$ plane for an effective pure spin model, as an analogue of (a) with full line representing $\sqrt{J_{2c}/J_1}$.}

\label{fig4}
\end{figure}

\section{Discussion and summary}

Results presented in Fig.~\ref{fig2} and Fig.~\ref{fig3} are typical
examples. In order to test the robustness of the results with respect
to charge fluctuations on the dots, we performed a detailed numerical
analysis summarized in the phase diagram in the parameter space
$(U/\Gamma,t_2/t_1)$, see Fig.~\ref{fig4}(a). The moment of the dot
$C$, $\langle\mathbf{S}_C^2\rangle$, is represented as a contour plot (dashed lines).
It is clear that for larger $\Gamma$ ({\it i.e.}, smaller $U/\Gamma$) the
moment is diminished which partially contributes to the
renormalization of $t_{2c}$.

The main reason for the renormalization can, in fact, be understood from an
effective spin model by tracing out the empty and doubly occupied states,
and representing the conduction band using a single spin $\mathbf{S}_D$ 
\cite{vojta,Wang, Mitchell,marcin,Luczak},
\begin{equation}\label{heisenberg}
{H}_{\text{eff}} = J_{1}
(\mathbf{S}_{A} + \mathbf{S}_B)\cdot\mathbf{S}_{C} + J_{2}
\mathbf{S}_{A}\cdot\mathbf{S}_{B}+ J_{D}
\mathbf{S}_{C}\cdot\mathbf{S}_{D},
\end{equation}
with the exchange couplings $J_{1,2}=4t^2_{1,2}/U$. 
First we consider a simple TQD system with $J_D=0$, {\it i.e.}, without spin
$\mathbf{S}_D$. The eigenstates are two doublets and a quadruplet (which always lies the highest in energy).
The ground states for the $z$-component of the total spin $S_{z}=1/2$ are given by
\begin{eqnarray}\label{d1}
|D_{1}\rangle&=&\frac{(|\uparrow_{A}\downarrow_{B}\rangle-|\downarrow_{A}\uparrow_{B}\rangle)\otimes|\uparrow_{C}\rangle}{\sqrt{2}},\\ \label{d2}
|D_{2}\rangle&=&\frac{2|\uparrow_{A}\uparrow_{B}\rangle\otimes|\downarrow_{C}\rangle-(|\uparrow_{A}\downarrow_{B}\rangle+|\downarrow_{A}\uparrow_{B}\rangle)\otimes|\uparrow_{C}\rangle}{\sqrt{6}}.\nonumber
\end{eqnarray}
The first state $|D_{1}\rangle$ is formed from the singlet state between the dots $A$ and $B$, for which the spin of the dot $C$ is decoupled from the other dot spins
\begin{equation}
\left\langle D_{1}|\mathbf{S}_C\cdot\mathbf{S}_A|D_{1}\right\rangle=\left\langle D_{1}|\mathbf{S}_C\cdot\mathbf{S}_B|D_{1}\right\rangle=0.
\end{equation}
The second state $|D_{2}\rangle$ is constructed from the triplet states between the dots $A$ and $B$ with $S_{z}=1$ and $0$, separately. The spin-spin correlation functions between the dots $A$ and $B$ for the above two states are 
\begin{equation}
\begin{split}
\left\langle D_{1}|\mathbf{S}_A\cdot\mathbf{S}_B|D_{1}\right\rangle&=-3/4, \\
\left\langle D_{2}| \mathbf{S}_A\cdot\mathbf{S}_B|D_{2}\right\rangle&=1/4.
\end{split}
\end{equation}
Therefore, for the states $|D_{1}\rangle$ and $|D_{2}\rangle$  the spin-spin correlations between the dots $A$ and $B$ are anti- and ferromagnetic, respectively. The eigenvalues are 
\begin{equation}
\begin{split}
E_{1}&=-12t^{2}_{1}/U, \\
E_{2}&=-4(4t^{2}_{1}-t^{2}_{2})/U.
\end{split}
\end{equation}
From the energy separation $E_{1}-E_{2}=J_1-J_2$ it is clear that $J_{2c}=J_1$ is the critical value separating entangled from unentangled states which explains results for large $U/\Gamma$.

Phase diagram presented in Fig.~\ref{fig4}(a) can qualitatively be
understood also in the regime of lower $U/\Gamma$ where $t_{2c}$ is
significantly lower that $t_1$. A simple analysis of the effective
Hamiltonian \eqref{heisenberg} for $J_D>0$, {\it i.e.}, for TQD coupled to
an additional singly occupied quantum dot $D$, reveals a crossing of
levels representing different ground states where the dots $A$ and $B$
form triplet or singlet configurations, if the interaction $J_D$ is
below or above some $J_{Dc}$, respectively. Critical value $J_{Dc}$
can be given analytically, 
\begin{equation}
J_{Dc}={(J_1-J_2)(J_1+2J_2) \over 2J_2}. 
\end{equation}
In Fig.~\ref{fig4}(b) we show the phase diagram $(J_1/J_D,\sqrt{J_2/J_1})$ where the full red line represents the separation between entangled and unentangled regimes. The topology of the dots is shown as an inset. 
Here $J_D/J_1$  plays the role of an effective Kondo coupling of the TQD to the leads, $J_D \propto \Gamma/U$. The dots $A$ and $B$ for small $J_2/J_1$ but large $J_D/J_1$ thus exhibit perfect entanglement when dots $C$ and $D$ are forced to couple into a singlet, which can be considered as an example of the entanglement monogamy concept \cite{monogamy}. However, due to the substantial renormalization of the local moment $\langle\mathbf{S}_C^2\rangle$ in the limit of small values, $U/\Gamma\lesssim4$, the application of the pure spin model is not fully justified there. This is the reason for different behavior of critical  $t_{2c}/t_1\propto U/\Gamma$ and  $\sqrt{J_{2c}/J_1}\propto \sqrt{J_1/J_D}$ in this regime.

In summary, the spin entanglement of electron pairs in a triangular
TQD, one dot being attached to non-interacting leads, is quantitatively analyzed in the regime of two competing many-body effects, namely the Kondo effect and the direct exchange interaction. In contrast to DQD configurations in which the Kondo effect suppresses the entanglement \cite{Ramsak2}, in this case, the Kondo effect -- in which a single spin in the dot C is screened by conduction electrons in an attached metallic lead - induces entanglement between the spins in the dots $A$ and $B$.  There are two regimes of interest, depending on the ratio of the interdot couplings $t_{2}/t_{1}$. On the one hand, for $t_{2}\gtrsim t_{1}$, spins in the dots $A$ and $B$ are always maximally entangled and the central dot C is in the Kondo regime with conduction leads, thus the entanglement is switched on due to the Kondo effect. For stronger interaction regime $U/\Gamma \gg 1$ this result is expected as a consequence of the quantum phase transition observed in TQD with triangular topology \cite{Wang,Mitchell}. Our analysis reveals that the switching between unentangled and perfectly entangled qubit pairs $A$ and $B$ persists also in the regime of  substantial charge fluctuations, either $\delta n_A^2$ or $\delta n_C^2$. In the regime of low $U/\Gamma$ there is a strong renormalization of $t_{2c}\ll t_1$, which can be understood within a simple analytically solvable pure spin model where the TQD is anti-ferromagnetically coupled to an additional singly occupied quantum dot. The induction of the singlet formation $C$-$D$ acts as switching of the entanglement between the dots $A$ and $B$ in a similar manner as the formation of the Kondo singlet in the original TQD system analyzed numerically.

\acknowledgments{Financial support from the EU FP7 project: Marie Curie ITN NanoCTM, the Slovenian Research Agency under contract no. P1-0044, National Science Centre (Poland) under the contract DEC-2012/05/B/ST3/03208, and computing facilities at J. Stefan Institute are gratefully acknowledged.}

\end{document}